\documentclass[conference]{IEEEtran}
\IEEEoverridecommandlockouts
\usepackage{spconf,amsmath,graphicx,cite,color,balance,amsfonts,subfigure,float}
\usepackage{stfloats}
\usepackage{cite}
\usepackage[T1]{fontenc}
\usepackage{stmaryrd}
\usepackage{amssymb}
\usepackage{mathrsfs}
\usepackage{amsfonts}  
\usepackage{algorithm}
\usepackage{algorithmic}
\usepackage{cuted}
\usepackage{balance}
\usepackage{tabularx}
\usepackage{booktabs}
\usepackage{stfloats}
\usepackage{multirow}
\usepackage{pgfplots}

\usepackage{dsfont}
\usepackage{esint}

\usepackage{url}

\usepackage{glossaries-extra}

\usepackage{enumitem}
\usepackage{tikz}
\usepackage{caption} 
\usepackage{tabularx}
\usepackage{comment}
\usepackage{relsize}\usepackage{subdepth}\allowdisplaybreaks

\usepackage{babel}

\vspace{-4.5em}
\title{Enhanced Automotive Radar Collaborative Sensing By Exploiting Constructive Interference}
\vspace{-1.2em}
\name{Lifan~Xu$^{\dagger}$,  Shunqiao~Sun$^{\dagger}$, and A.~Lee~Swindlehurst$^{\ddagger}$ 
\thanks{This work was supported in part by U.S. National Science
Foundation (NSF) under Grants CCF-2153386 and ECCS-2340029.}
\vspace{-0.5em} 
}
\address{ \hspace{-0.7em} \hspace{-0.7em} 
 \hspace{-0.9em}  $^{\dagger}$Department of Electrical and Computer Engineering, University of Alabama, Tuscaloosa, AL 35487\\
 \hspace{-0.9em}  $^{\ddagger}$Center for Pervasive Communications and Computing, University of California, Irvine, CA 92697
}

\begin{document}

  \ninept
\maketitle
\begin{abstract}
Automotive radar emerges as a crucial sensor for autonomous vehicle perception. As more cars are equipped radars, radar interference is an unavoidable challenge. Unlike conventional approaches such as interference mitigation and interference-avoiding technologies, this paper introduces an innovative collaborative sensing scheme with multiple automotive radars that exploits constructive interference. Through collaborative sensing, our method optimally aligns cross-path interference signals from other radars with another radar's self-echo signals, thereby significantly augmenting its target detection capabilities. This approach alleviates the need for extensive raw data sharing between collaborating radars. Instead, only an  optimized weighting matrix needs to be exchanged between the radars. This approach considerably decreases the data bandwidth requirements for the wireless channel, making it a more feasible and practical solution for automotive radar collaboration. Numerical results demonstrate the effectiveness of the constructive interference approach for enhanced object detection capability. 
\end{abstract}


\section{Introduction}
\label{sec:intro}

In the era of autonomous driving, vehicles rely on a sophisticated sensor array, including cameras, lidar, ultrasonic sensors, and radar, to ensure a comprehensive and intelligent driving experience. Automotive radar stands out as a linchpin, offering commendable attributes such as cost-effectiveness, minimal processing requirements, and adaptability to various weather conditions.
Playing a pivotal role in autonomous vehicles, automotive radar facilitates essential advanced driver assistance features like automatic cruise control, blind-spot detection, automatic emergency braking, and autonomous driving, making it an integral component for enhancing overall driving safety and efficiency\cite{SUN_SPM_Feature_Article_2020,engels2021automotive,waldschmidt2021automotive}.
Despite its benefits, the increasing prevalence of vehicles equipped with automotive radars, some with multiple radar sensors, introduces a notable challenge, i.e., mutual interference between these radars. Without effective interference mitigation, automotive radar sensors experience performance degradation, posing an urgent problem that demands prompt solutions to ensure the seamless operation of advanced autonomous systems \cite{alland2019interference,aydogdu2020radar,jin2021fmcw,li2024performance}.

Various signal processing techniques are employed to mitigate interference, with several successfully implemented in automotive radar systems. In conventional frequency modulated continuous wave (FMCW) radars, diverse methods have been utilized to rectify samples at the receiving end that may have been disturbed by interfering sources or to reconstruct an uninterrupted signal, including sparse sampling\cite{bechter2017automotive,uysal2018mitigation}, adaptive beamforming\cite{Rameez_Radarconf_2018}, and principle singular value decomposition methods\cite{sun2020removing}. Beyond these computation-intensive techniques, the gating method provides a straightforward and efficient alternative for automotive radar, utilizing simple and effective threshold filtering\cite{SUN_SPM_Feature_Article_2020}.
On the transmitter side, the goal is to avoid or minimize interference to other radars, achieved through well-designed waveforms or strategic adjustments of the operational frequency or sparse pulsing \cite{bechter2016bats,sun20214d,xu2023automotive}. Compared with FMCW radar, phase-modulated continuous wave (PMCW) radar can significantly mitigates mutual radar interference with orthogonal codes that are already designed for communications \cite{overdevest2019uncorrelated,xu2019doppler,bourdoux2021pmcw}. Regarding the mutual interference, the state-of-the-art  algorithms focus on mutual interference identification, suppression or mitigation, within a stand alone automotive radar \cite{alland2019interference}.

Multiple automotive radars could cooperate with each other to 
enhance detection performance\cite{feger201277,edstaller2021cooperative,Sun_collaborative_sensing_EuRAD_2023}. For the coherent combination of signals from different radars, precise time and phase synchronization among individual nodes is crucial. To address this synchronization issue, centralized data processing methods are often employed\cite{feger201277,gottinger2021coherent}. For example, a system with cooperative radars synchronized with centralized real-time computation and fusion with data from other time-synchronized sensors is proposed in \cite{gottinger2021coherent} to increase angluar resolution. Decentralized technology utilizes a reference target for data calibration and offset compensation, eliminating unknown frequency and phase errors in the beat spectrum of each radar due to noncoherent operation. This approach also reduces the computational cost compared to the centralized approach\cite{edstaller2021cooperative}. Additionally, the global positioning system (GPS) and GPS-disciplined oscillator (GPSDO), have been developed to precisely synchronize multiple radars\cite{sandenbergh2017synchronizing}.

In this paper, we propose a collaborative sensing scheme with multiple automotive PMCW radars that exploits constructive interference. The key idea is to transform the interference into useful information to enhance target detection performance. In our proposal, we assume the radars are time-synchronized and the radar channel state information \cite{JRC_80211ad_TVT_2018} can be estimated beforehand, and it remains relatively stable across consecutive coherent processing intervals (CPIs). By designing and applying a weighting matrix to the waveform of the cooperative radar, the cross-path signal can be effectively aligned with the self-echo signal of another radar to boost its detection performance. Unlike conventional multistatic radar systems, which require extensive data sharing for coherent processing and thus consume more resources \cite{feres2023over,moussa2024multi}, the proposed method only shares the optimized weighting factors rather than the raw measurement. Numerical results demonstrate the effectiveness of the proposed collaborative sensing scheme that exploits constructive interference.

\section{System Model}
\label{problem formulation}

\begin{figure} 
\centering
{\includegraphics[width=1\columnwidth]{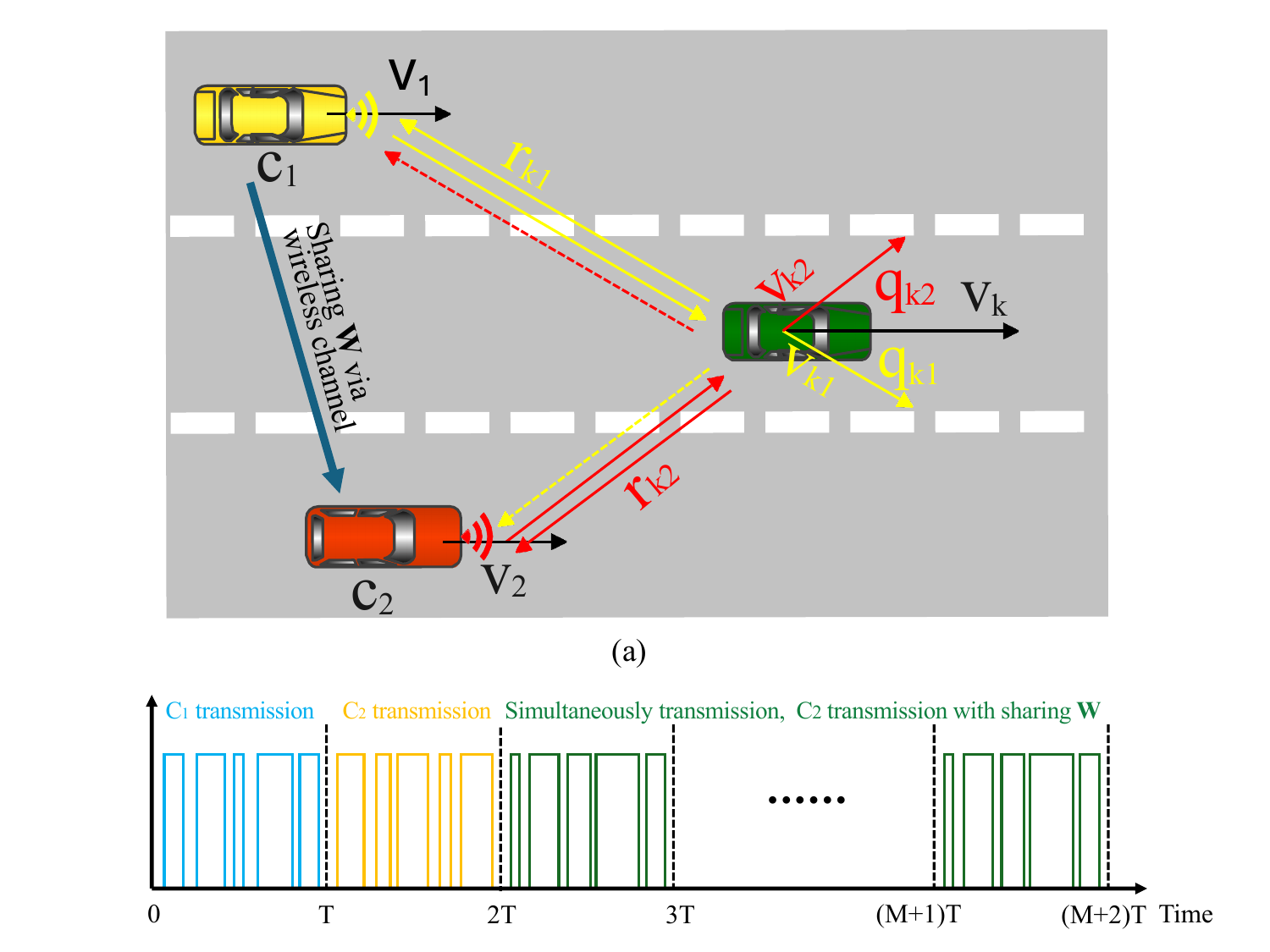}}
\caption{ (a) System model of multiple automotive radars exploiting constructive interference. The solid and dashed lines denote self-echoes and cross-path echoes, respectively; (b) The unilateral enhancement transmission scheme.}
\label{fig_model}
\vspace{-3mm}
\end{figure}

In this paper, we consider two single-input-single-output (SISO) automotive radars mounted on two vehicles, denoted as $c_1$ and $c_2$, shown in Fig. \ref{fig_model}. These vehicles are moving with velocities $v_1$ and $v_2$, respectively. Assume there are $K$ objects in the field of views (FOVs) of both radars, while the host vehicles are not in the FOV of each other. For simple illustration, a moving vehicle with velocity $v_k$ is shown in Fig. \ref{fig_model}.  Without loss of generality, we asumme both host vehicles and target vehicle are moving along the same direction. The target's relative radial velocity observed at radar $c_1$ and radar $c_2$ are $v_{k1}= (v_k- v_1){\rm cos}\theta_{k1}$ and $v_{k2}= (v_k- v_2){\rm cos}\theta_{k2}$, where $\theta_{k1}$ and $\theta_{k2}$ are the angle between the direction of motion of the target vehicle and the radial direction w.r.t. radar $c_1$ and radar $c_2$, respectively. We assume each radar is not in the FOV of the other one, and thus direct paths between radars are not considered.

The radars transmit phase-modulated continuous wave (PMCW) signals with unimodular code sequences $\bf c_1$ and $\bf c_2$, both of which are derived from the same code family with code length $N$, and are known by both radars. These sequences are used to modulate the carrier frequency over a short duration for target detection. 
For each pulse duration time $T$, there are $N$ subpulses (chips) transmitted and each chip has time duration $T_c = T/N$. To avoid mutual interference among the radars, the code set is desired to have the following ideal cross-correlation properties.
\begin{align}
    \mathbf{R}_{\mathbf{c}_1\mathbf{c}_2}[l] = \sum_{n=1}^N \mathbf{c}_1[n]\mathbf{c}^*_2[n-l] = 0, \;\forall \; l  = 1,2,\cdots,N_1. \label{eq_ideal_cross_corr}
\end{align}
The time lag $l$ implies a relative time shift of the two sequences. The zero cross-correlation length $N_1$ is limited by the maximum delay difference \cite{he2009designing}. 
To estimate Doppler, $M$ pulses are transmitted in one CPI.

\subsection{Multiple PMCW Radars Under Mutual Interference}
\label{sec_mutual_interference}
As shown in Fig. \ref{fig_model}, the received signal at each radar contains two components that are reflected from the target, i.e., a self-echo from the direct path and a cross-echo from the cross-path. Without collaboration, each radar would treat the cross-path signals as mutual interference.

The transmit signal of the radar $c_1$ is given by
\begin{align}
    x_{c_1}(m,t) & = \sum_{n = 1}^N \mathbf{c}_1[n]{\rm rect}\left\{\frac{t-nT_c-mT}{T_c}  \right\} e^{{\rm j}2\pi f_c(t-mT)}\nonumber \\ 
    &= s_{c_1}(t-mT)e^{{\rm j}2\pi f_c(t-mT)},
\end{align}
where $s_{c_1}(t-mT) = \sum_{n = 1}^N \mathbf{c}_1[n]{\rm rect}\left\{\frac{t-nT_c-mT}{T_c}  \right\}$ is the phase modulated signal. Here the carrier frequency is $f_c$, and the rectangular pulse window function is defined as
\begin{align}
{\rm{rect}}\left( {\frac{{t - \tau }}{T}} \right) = \left\{ \begin{array}{l}
 1, \quad\tau   \le t \le \tau  + T, \\ 
 0, \quad {\rm{otherwise}} .\\ 
 \end{array} \right.
\end{align}

The noise-free received self-echo signal ${y_s(m,t)}$ at radar $c_1$ is a weighted delayed version of ${x_{c_1}(m,t)}$, expressed as
  \begin{align}
     y_s(m,t) = &\sum_{k=1}^{K}\alpha_{k}x_{c_1}(m,t-\tau_k)\nonumber\\
     =&\sum_{k=1}^{K}\alpha_{k}s_{c_1}(t-mT-\tau_k)e^{{\rm j}2\pi f_c (t-mT-\tau_k)},
 \end{align}
where the round-trip transmission delay for the $k$-th target is ${\tau_{k} = 2(r_{k1}+v_{k1}t)/c}$ with ${v_{k1}} = \left( {{v_k} - {v_1}} \right)\cos \left( {{\theta _{k1}}} \right)$. Therefore, the received self-echo signal can be expressed as
\begin{align}
   &y_s(m,t) =  \sum_{k=1}^{K}\alpha_{k}s_{c_1}\big(t-mT-\frac{2(r_{k1}+v_{k1}t)}{c}\big)\nonumber\\
  &\quad \qquad \qquad \qquad \cdot e^{{\rm j}2\pi f_c \big(t-mT-\frac{2(r_{k1}+v_{k1}t)}{c}\big)}\nonumber\\
  &\approx \sum_{k=1}^{K}\hat{\alpha}_{k}s_{c_1}\big(t-mT-\frac{2r_{k1}}{c}\big)e^{{\rm j}2\pi f_c(t-mT)}e^{{-\rm j2}\pi f_d^{k1}t},
\end{align}
During a pulse, the term $2v_{k1}t/c $ in $s_{c_1}(t-mT-\frac{2r_{k1}}{c})$ is omitted since we assume $2v_{k1}t/c  \ll 2r_{k1}/c$. The phase term $ 2f_cr_{k1}/c$ does not change in fast time and is only associated with the range, so it can be absorbed into the reflection coefficient. The Doppler frequency of the $k$-th target is $f_d^{k1} = 2v_{k1}f_c/c$.

The received self-echo signal $y_s(m,t)$ is demodualated through mixing with the conjugate of the carrier waveform, and therefore 
\begin{align}
     \label{mixed-signal}
    \hat {y}_s(m,t) =  \sum_{k=1}^{K}\hat{\alpha}_{k}s_{c_1}\big(t-mT-\frac{2r_{k1}}{c}\big)e^{{-\rm j2}\pi f_d^{k1}(t-mT)} .
\end{align}
The received data is sampled with a sampling rate of $1/T_c$. The sampled receive signal is given by
\begin{align}
    &\hat{{y}}_s(m,n)\nonumber\\
    = &\sum_{k=1}^{K}\hat{\alpha}_{k}s_{c_1}\big(nT_c-mT-\frac{2r_{k1}}{c}\big)e^{{-\rm j2}\pi f_d^{k1}(nT_c-mT)} \nonumber\\
    = &\sum_{k=1}^{K}\hat{\alpha}_{k}\mathbf{c}_1[n-\hat{n}_{r_{k1}}]e^{{-\rm j2}\pi f_d^{k1}nT_c}e^{{\rm j}2\pi f_d^{k1} mT},
\end{align}
where $\hat{n}_{r_{k1}} = \left\lceil \frac{2r_{k1}}{Tc} \right\rceil $ is the code shift w.r.t. the $k$-th target.

In the assumption of an equivalent number of targets within the field of view of both radar systems $c_1$ and $c_2$, the sampled cross-path echo $\hat{y}_{c2}(m,n)$ concurrently received by radar $c_1$ from radar $c_2$ is as follows
\begin{align}
    \hat{{y}}_{c2}(m,n)   
    = &\sum_{k=1}^{K}\bar{\alpha}_k\mathbf{c}_2[n-\hat{n}_{r_{k12}}]e^{{-\rm j2}\pi f_d^{\hat{k}}nT_c}e^{{\rm j}2\pi f_d^{\hat{k}} mT},
\end{align}
where $\bar{\alpha}_{k}$ is obtained from $\alpha_k$ by absorbing the constant phase $ 2f_cr_{k12}/c$, and $\hat{n}_{r_{k12}} = \left\lceil {r_{k12}}/{T_c} \right\rceil $ is the $\mathbf{c}_2$ code shift due to the location of $k$-th target, where $r_{k12} = r_{k1}+r_{k2}$ is the distance combining transmission range from radar $c_2$ to the target and from the target to radar $c_1$, and $f_d^{\hat{k}}= \left[ {2{v_k} - \left( {{v_1} + {v_2}} \right)} \right]\cos \left( {{\theta _{k1}}} \right)f_c/c$.
As a result, the echo received by radar $c_1$ from the target comprises two components, i.e., the self-echo and the cross-path signal from radar $c_2$. Therefore, the sampled received signal for radar $c_1$  can be expressed as follows 
\begin{align}
     \hat{\mathbf{Y}}_1  = {\bf{Y}}_s^1 + {\bf{Y}}_{c2}^1 + {\bf E}_1, \label{eq_received_signal}
\end{align}
where $ {\bf{Y}}_s^1 = \mathbf{C}_{K1}\odot\mathbf{D}_{K1}\mathbf{\Sigma}\mathbf{B}^T_{K1},{\bf{Y}}_{c2}^1 = \mathbf{C}^{'}_{K2}\odot\mathbf{D}_{\hat{K}}\mathbf{\Sigma}^{'}\mathbf{B}^T_{\hat{K}}$ are the self-echo and cross-path signals, respectively. Here,   ${\bf E}_1$ is random noise, and $\odot$ is the element-wise product operator, and $\mathbf{C}_{K1} = \left[ \mathbf{c}_1[n-\hat{n}_{r_{11}}],\cdots,\mathbf{c}_1[n-\hat{n}_{r_{K1}}] \right] \in \mathbb{R}^{N\times K}$ is the shifted code  matrix, and $\mathbf{D}_{K1} = \left[ \mathbf{d}_{11},\mathbf{d}_{21},\cdots,\mathbf{d}_{K1} \right]\in \mathbb{C}^{N\times K} $ with $\mathbf{d}_{k1} = \left[e^{{-\rm j2}\pi f_d^{k1}T_c},e^{{-\rm j2}\pi f_d^{k1}2T_c},\dots,e^{{-\rm j2}\pi f_d^{k1}(N-1)T_c} \right]^T$. The reflect coefficients are ${\bf{\Sigma }} = {\rm{diag}}\left( {\left[ {{\alpha _1}, {\alpha _2},\cdots ,{\alpha _K}} \right]} \right)$. 
Here, the Doppler manifold is ${\bf{B}}_{K1} = \left[ {{\bf{b}}_1( {f_d^{11}} ), {\bf{b}}_2( {f_d^{21}} ), \cdots,{\bf{b}}_K( {f_d^{K1}} )} \right] \in {\mathbb C}^{M\times K}$ with ${\bf{b}}_k( {f_d^{k1}} ) = {[ {{e^{\mathrm{j}2\pi f_d^k{T}}}, {e^{\mathrm{j}2\pi f_d^k{2T}}}, \cdots ,{e^{\mathrm{j}2\pi f_d^kM{T}}}} ]^T}$.  Similarly, matrices $\mathbf{C}^{'}_{K2}$,
$\mathbf{D}_{\hat{K}}$, $\mathbf{B}_{\hat{K}}$ and  $\mathbf{\Sigma}^{'}$ are  shifted transmit code, Doppler manifolds in fast-time, slow-time, reflect coefficients in cross-path.

After applying correlation in the fast-time domain for the range profile corresponding to radar $c_1$, we have
\begin{align}
    {r}_1^1(n,m) &= {\rm corr}  \left[\hat{{\rm \mathbf{Y}}}_1(n,m), \mathbf{c}_1[n-\hat{n}_{r_{k1}}] \right] \nonumber\\
    & = \sum_{k=1}^{K}\hat{\alpha}_{k}\delta(n-\hat{n}_{r_{k1}})e^{{\rm j}2\pi f_d^{k1} mT} + \xi_1,
\end{align}
where  ${\xi _1} = {\rm{corr}}\left[ {{\bf{Y}}_{c2}^1\left( {n,m} \right) + {{\bf{E}}_1}\left( {n,m} \right),{{\bf{c}}_1}\left[ {n - {{\hat n}_{{r_{k1}}}}} \right]} \right]$ and $\delta(\cdot)$ is the Dirichlet function. 
Here, we assume the relative velocities of the targets are small and result in small Doppler shifts. Thus the additive frequency shifts $e^{{-\rm j2}\pi f_d^{k1}nT_c}$ in the fast-time domain can be omitted without compensation. The correlation result ${\xi _1}$ corresponding to cross-path signal is around noise floor for ideal zero cross-correlation code set defined in equation (\ref{eq_ideal_cross_corr}).

Then, discrete Fourier transform (DFT) is applied along slow-time domain to obtain the range-Doppler estimation, i.e., 
\begin{align}
   & R^1_1(n,\omega) = \sum_{m=1}^{M}r_1^1(n,m)e^{{\rm -j}2\pi \frac{\omega}{M}m}\nonumber\\
    & =  \sum_{k=1}^{K} \hat{\alpha}_{k}\delta(n-\hat{n}_{r_{k1}})e^{{\rm -j}\pi(M-1)(\frac{\omega}{M}-f_d^{k1}) }\rho(\omega) +\Xi_1, \label{eq_Doppler_DFT}
\end{align}
where $\rho(\omega) = \frac{{\rm sin}[\pi M(\frac{\omega}{M}-f_d^{k1}T)]}{{\rm sin}[\pi(\frac{\omega}{M}-f_d^{k1}T)]}$ and the DFT of the noise is denoted as $\Xi_1$. In a similar way, the range-Doppler estimation at radar $c_2$ can be obtained.

\subsection{Collaborative Sensing with Constructive Interference}

We propose a collaborative sensing scheme, as illustrated in Fig. \ref{fig_model}, aiming to boost the detection performance of radar $c_1$. The received power of the self-echo at radar $c_1$ is $P_1 \propto 1/r_{k1}^4$, where $r_{k1}$ is the range w.r.t. radar $c_1$. The cross path power at radar $c_1$, i.e., the bistatic radar path, is $P_{21} \propto 1/(r_{k2}^2r_{k1}^2)$, where $r_{k2}$ is the range w.r.t. radar $c_2$. When $r_{k2} < r_{k1}$, it is possible that the power of cross-path signal at radar $c_1$ is higher than its self-echo signal, and the cross-path signal can be utilized constructively. The multiple radars are assumed to be time-synchronized, and their codebooks are known a priori by each radar. In contrast to the aforementioned non-cooperative approach, where each radar treats the cross-path signals as interference, multiple automotive radars can collaborate to form a multistatic radar system, thereby enhancing overall sensing performance. However, the system model proposed in this paper distinguishes itself from traditional multistatic radar by eliminating the need for each radar to share its raw measurements for coherent processing.

As shown in Fig. \ref{fig_model} (b), in the first pulse, only $c_1$ transmits with code $\mathbf{c}_2$ and receives the self-echo signal $\mathbf{y}^1_s = \mathbf{C}_{K2}\odot\mathbf{D}_{K1}\mathbf{\Sigma}\mathbf{b}^{1}_{K1}$, where $\mathbf{b}^{1}_{K1} = [f_d^{11},f_d^{21},\cdots ,f_d^{K1}]^T $. In the subsequent pulse, only $c_2$  transmits with code $\mathbf{c}_2$,  and radar $c_1$ receives the cross-path signal  $\mathbf{y}^1_{c2} = \mathbf{C}^{'}_{K2}\odot\mathbf{D}_{\hat{K}}\mathbf{\Sigma}^{'}\mathbf{b}^{1}_{\hat{K}}$ with $\mathbf{b}^{1}_{\hat{K}} = [f_d^{\hat{1}},f_d^{\hat{2}},\cdots ,f_d^{\hat{K}}]^T$. 
In this paper, we assume the radar channel state information w.r.t. the cross-path, such as range, Doppler and reflect coefficients, has been estimated in previous CPI, and it does not change significantly in successive CPIs.

To use the cross-path interference signal  $\mathbf{y}^1_{c2}$ in a constructive way to enhance the target detection capability for radar $c_1$, we introduce the weighting matrix ${\bf W} \in {\mathbb C}^{N\times N}$ that is applied to the transmit code ${\bf c}_2$ at the transmitter of radar $c_2$ to align the new cross-path echo from $c_2$, i.e., ${\bf{y}}_{c2}^{1, {\rm{new}}} = \left(\mathbf{W}\mathbf{C}^{'}_{K2}\right)\odot\mathbf{D}_{\hat{K}}\mathbf{\Sigma}^{'}\mathbf{b}^1_{\hat{K}}$, with the self-echo of $c_1$, i.e., $\mathbf{y}^1_s$. 
Thus it is desired to align the cross-path signal with self-echo signal in phase, i.e.,
\begin{align}
\angle \left(\mathbf{C}_{K2}\odot\mathbf{D}_{K1}\mathbf{\Sigma}\mathbf{b}^1_{K1} \right) =
\angle \left( \left(\mathbf{W}\mathbf{C}^{'}_{K2}\right)\odot\mathbf{D}_{\hat{K}}\mathbf{\Sigma}^{'}\mathbf{b}^1_{\hat{K}} \right),
\end{align}
where $\angle(\cdot)$ denotes the element-wise phase operator. Once the phases are aligned, the superposition of self-echo and cross-path signals at radar $c_1$ will  add up constructively.
Intuitively, if the phases are aligned, the imaginary part of the product between the self-echo and the complex conjugate of the cross-path signal components is expected to be zero. Let $\mathbf{u} = \mathbf{y}^1_s\odot (\mathbf{y}^{1, {\rm new}}_{c2})^*$. Then, the phase alignment optimization problem is formulated as
\begin{align}
{\text{minimize}}& \quad
\parallel \Im(\mathbf{u}) \parallel_2 \nonumber \\
 \text{subject to}
 &  \quad \; {\Re}\left( {\bf{u}} \right) =  \left| {{\bf{y}}_s^1} \right| \cdot \left| {{\bf{y}}_{c2}^{1,{\rm new}}} \right|, \label{eq_W_opt}
\end{align} 
where 
$\Re(\cdot)$ and $\Im(\cdot)$ represent the real and imaginary parts, respectively. The conjugate operator is $(\cdot)^{*}$ and $\left|  \cdot  \right|$ is element-wise absolute value operator. Here, the  constraint is to make sure the new transmit code after applying weight $\bf W$ at radar $c_2$ is still unimodular. In other words, the signal-to-noise ratio (SNR) gain in radar $c_1$ by exploiting constructive interference is attributed solely to the phase alignment and not to any increase in the transmitted power of radar $c_2$. Consequently, the CVX toolbox can be employed to solve this problem.
Once the matrix $\bf W$ is obtained, it is shared to radar $c_2$ via a wireless channel, and $\bf W$ remains valid for the duration of one CPI.

From the third pulse, both radars transmit simultaneously with radar $c_1$ transmitting code ${\bf c}_1$ and radar $c_2$ transmitting ${\bf Wc}_2$. After $M$ pulses, the sampled received signal at radar $c_1$ is
\begin{align}
    \bar{\mathbf{Y}}_1  = {\bf Y}_s^1 + {\bf Y}_{c2}^{1, {\rm new}} + {\bf E}_1, \label{eq_received_signal_collaborative}
\end{align}
where ${\bf Y}_s^1  = \mathbf{C}_{K1}\odot\mathbf{D}_{K1}\mathbf{\Sigma}\mathbf{B}^T_{K1}$ is self-echo signal, ${\bf Y}_{c2}^{1, {\rm new}} = (\mathbf{W}\mathbf{C}^{'}_{K2})\odot\mathbf{D}_{\hat{K}}\mathbf{\Sigma}^{'}\mathbf{B}^T_{\hat{K}} $ is well aligned cross-path signal and ${\bf E}_1$ is random noise. Utilizing the waveform orthogonality defined in equation (\ref{eq_ideal_cross_corr}),
and applying correlation in the fast-time domain on measurement at radar $c_1$ with code ${\bf c}_1$, we have the range profile corresponding to the self-echo signal  as
\begin{align}
    \bar{r}_1^1(n,m) &= {\rm corr}  \left[\bar{{\rm \mathbf{Y}}}_1(n,m), \mathbf{c}_1[n-\hat{n}_{r_{k1}}] \right] \nonumber\\
    & = \sum_{k=1}^{K}\hat{\alpha}_{k}\delta(n-\hat{n}_{r_{k1}})e^{{\rm j}2\pi f_d^{k1} mT} + \xi^{'}_1.  
\end{align}
Here,  $\xi _1^{'} = {\rm{corr}}\left[ {{\bf{Y}}_{c2}^{1,{\rm{new}}}\left( {n,m} \right) + {{\bf{E}}_1}\left( {n,m} \right),{{\bf{c}}_1}\left[ {n - {{\hat n}_{{r_{k1}}}}} \right]} \right]$ is at noise floor since the cross-path signal is aligned to self-echo signal when ${\bf c}_2$ is transmitted at radar $c_1$. 

Similarly, applying correlation in the fast-time domain on measurement at radar $c_1$ with code ${\bf c}_2$,  we have the range profile corresponding to the aligned cross-path signal as
\begin{align}
        \bar{r}_2^1(n,m) &= {\rm corr}  \left[{\rm }\bar{{\rm \mathbf{Y}}}_1(n,m), \mathbf{c}_2[n-\hat{n}_{r_{k1}}] \right] \nonumber\\
        & = \sum_{k=1}^{K}\bar{\alpha}_{k}\delta(n-\hat{n}_{r_{k1}})e^{{\rm j}2\pi f_d^{k1} mT} + \xi^{'}_2.
\end{align}
Here, $\xi _2^{'} = {\rm{corr}}\left[ {{\bf{Y}}_s^1\left( {n,m} \right),{{\bf{c}}_2}\left[ {n - {{\hat n}_{{r_{k1}}}}} \right]} \right]$ is close to zero when perfect orthogonal codes are utilized. With the proposed alignment, the range profiles corresponding to cross-path $ \bar{r}_2^1(n,m)$ is no longer interference  under the non-cooperative scheme defined in Section \ref{sec_mutual_interference}. Further, the indices of  range peaks in $ \bar{r}_2^1(n,m)$ are identical to those in $\bar{r}_1^1(n,m)$ corresponding to self-echo. 

Apply the same DFT operation defined in (\ref{eq_Doppler_DFT}) on $\bar{r}_1^1(n,m)$ and $\bar{r}_2^1(n,m)$, the range-Doppler profiles corresponding to both self-echo signal $\bar{R}^1_1(n,\omega)$ and cross-path signal $\bar{R}^1_2(n,\omega)$ are obtained, respectively. The final enhanced range-Doppler profile is obtained by non-coherently adding the two range-Doppler profiles together.
\begin{align}
    R_{\rm enhanced}^1(n,\omega) &=\bar{R}^1_1(n,\omega)+\bar{R}^2_1(n,\omega).
\end{align}
With constructive interference, the peaks in the enhanced range-Doppler profile exhibit increased amplitudes, leading to a notable improvement in the detectability of the targets.  The constructive interference method is summarized in Algorithm \ref{algo1}.

\begin{algorithm}
	\caption{Collaborative Sensing via Constructive Interference} 
	\begin{algorithmic}[1] \label{algo1}
        \STATE {\bf Initialization:} radars $c_1$ and $c_2$ are synchronized, and their code sets $\mathbf{c}_1$ and $\mathbf{c}_2$ are known by each other.  
        \STATE {\bf Coordinate Phase:} In the first pulse, radar $c_1$ transmits with $\mathbf{c}_2$ and collects self-echo signal $\mathbf{y}^1_s$. In the second pulse, radar $c_2$ transmits with $\mathbf{c}_2$ and radar $c_1$ collects cross-path signal $\mathbf{y}^1_{c2}$. Obtain the weighting matrix $\bf W$ by solving optimization problem (\ref{eq_W_opt}) at $c_1$ and share it with $c_2$ via a wireless channel.
        \STATE {\bf Simultaneously Transmission:} Starting from the third pulse, $c_1$ transmits with code $\mathbf{c}_1$ and $c_2$ transmits with code $\bf Wc_2$.
        \STATE {\bf Enhanced Range-Doppler Profiles:} Correlate the received signal at radar $c_1$ with $\mathbf{c}_1$ and $\mathbf{c}_2$ at $c_1$. Apply DFT to obtain the range-Doppler profiles corresponding to self-echo $\bar{R}^1_1(n,\omega)$  and cross-path $\bar{R}^1_2(n,\omega)$, respectively. Add up the two range-Doppler profiles non-coherently.   
	\end{algorithmic} 
\end{algorithm}

{\bf Discussion:} The coordination phase, which lasts for the duration of two pulses, constitutes a relatively small portion of the overall CPI that is usually consisting of hundreds of pulses. Moreover, the weighting matrix $\bf W$ can be designed to align the cross-path signal with the self-echo signal corresponding to the code ${\bf c}_1$ transmitted by radar $c_1$. Under this alignment, the received signal at radar $c_1$ would combine coherently in time domain. However, the gain in the range-Doppler profile is limited because the orthogonality in the code domain is not  exploited.

\section{Numerical Results}
We consider a collaborative automotive radar network consisting of two PMCW automotive radars with random binary phase codes of length $N = 256$. The carrier frequency is $f_c = 77$ GHz, and the bandwidth is $B = 120$ MHz. The ranges of a point target w.r.t. radars $c_1$ and $c_2$ are $r_1 = 5$ m and $r_2 = 10$ m. The relative velocity of the target to two radars is $v= 10 $ m/s. The SNR of received signals at $c_1$ and $c_2$ are $10$ dB.

\begin{figure}[ht]
\vspace{-4mm}
\centering
\subfigure[]{\includegraphics[height=1.2 in]{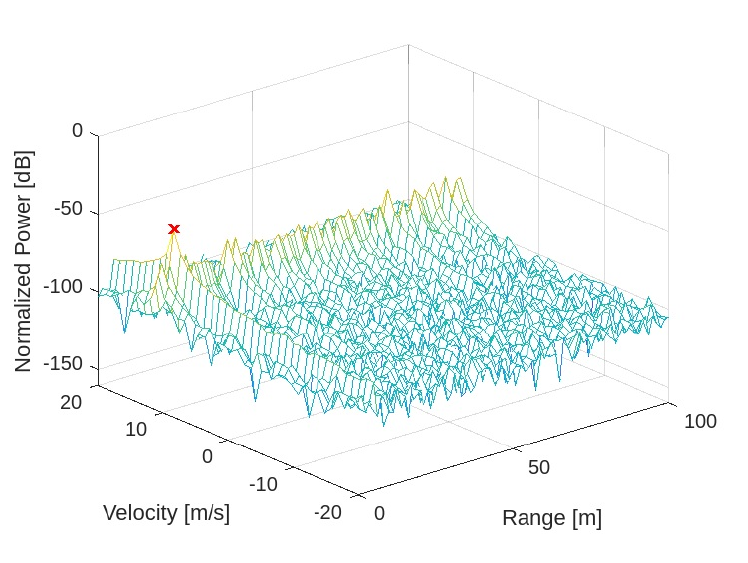}}
\subfigure[]{\includegraphics[height=1.2 in]{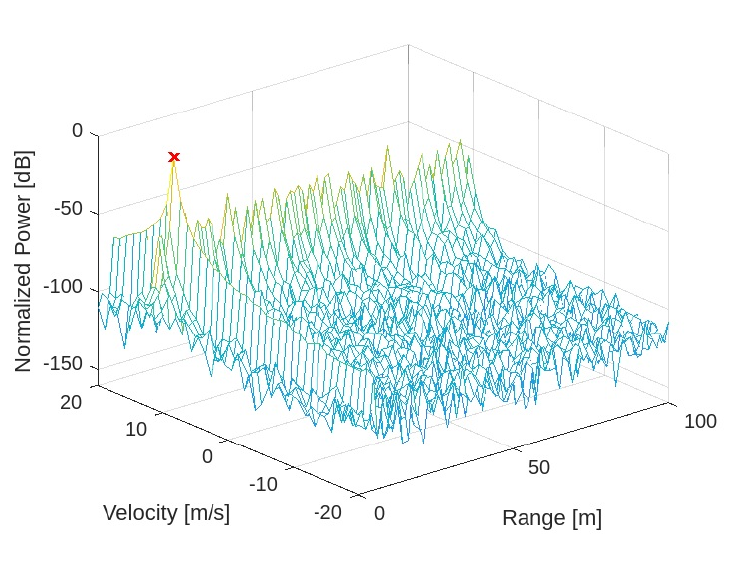}} 
\subfigure[]{\includegraphics[height=1.2 in]{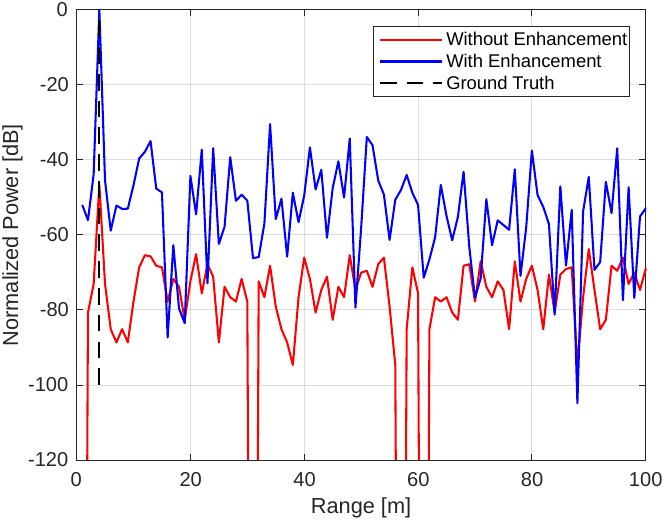}} 
\vspace{-2mm}
\subfigure[]{\includegraphics[height=1.2 in]{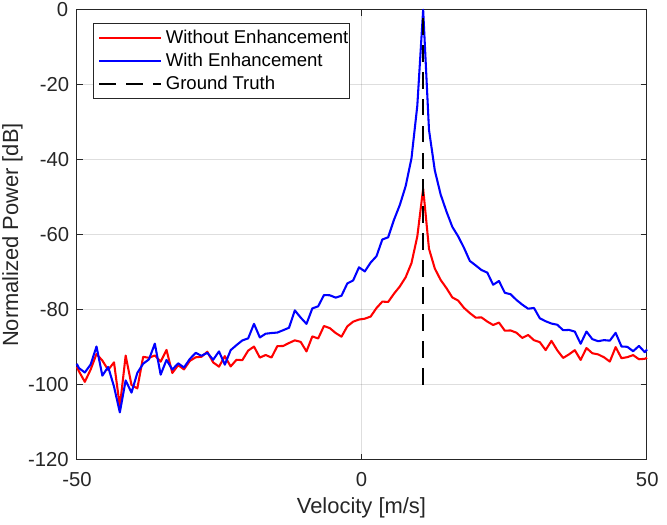}}
\subfigure[]{\includegraphics[height=1.2 in]{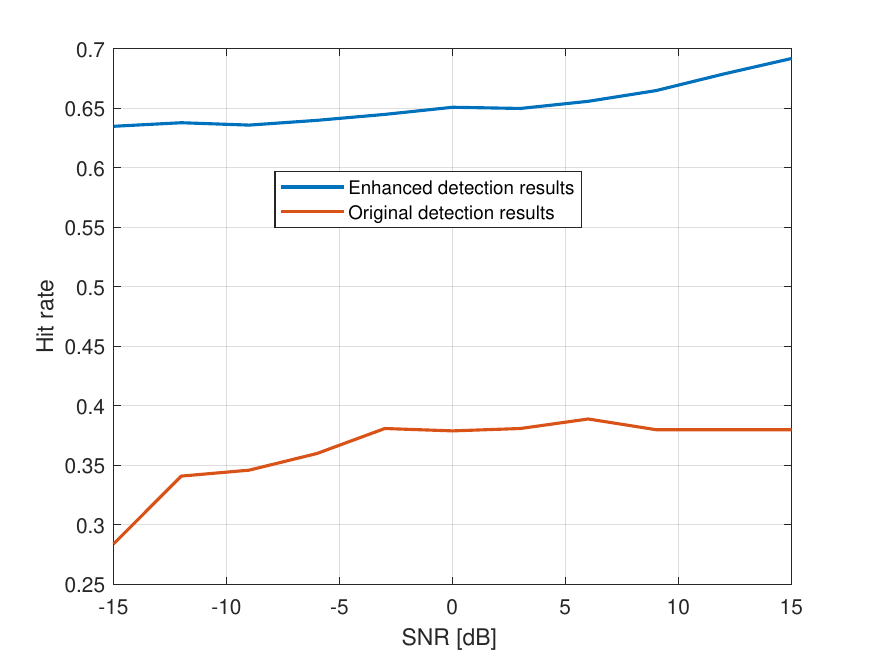}}
\vspace{-2mm}
\caption{Performance comparison for radar $c_1$: (a) Range-Doppler spectrum without enhancement; (b) Range-Doppler spectrum with enhancement;  (c) Range estimation cut; (d) Doppler estimation cut; (e) Hit rate of detection.} 
\label{Range-Doppler}
\vspace{-6mm}
 \end{figure}

For radar $c_1$, as illustrated in Fig. \ref{Range-Doppler} (a), the original range-Doppler spectrum exhibits higher sidelobes along the range dimension due to the nonideal orthogonal code set. By employing optimization with cross-path information, as depicted in Fig. \ref{Range-Doppler}
(b), the true target becomes more distinct compared to the original representation. Figures \ref{Range-Doppler} (c) and (d) plot the range and Doppler spectrum cut with and without enhancement. In this analysis, the signals without enhancement are normalized relative to the enhanced signals to effectively demonstrate the performance improvements attributable to constructive interference. With optimization enhancement, the peak sidelobe level (PSL) of the range profiles decreases by about $20$ dB.

To statistically assess the detection enhancement performance facilitated by constructive interference, we apply the hit-or-miss criterion \cite{na2018tendsur}. This criterion is employed to evaluate the range-Doppler recovery rate under various input SNR values. In this context, a ``hit'' implies that the absolute error of the detected targets falls within the range resolution.

We placed two targets with normalized reflection coefficients, where $\alpha_1=1$ and $\alpha_2$ are randomly drawn from the interval $\left [ 0.2, 0.5 \right ]$. The range of these two targets is uniformly selected at random from $\left [ 10, 100 \right ]$ meters. For each input SNR, chosen from 11 uniformly spaced values in the interval $\left[-15, 15\right]$ dB, we conduct $1,000$ Monte Carlo simulations.
As shown in Fig. \ref{Range-Doppler} (e), the enhanced detection exhibits a hit rate that is nearly double that of the conventional detection method for all SNRs.

\section{Conclusions}
We introduced a collaborative automotive radar sensing approach that constructively leverages mutual interference. By optimizing cross-path signals to align with self-echo signals and the effective utilization of orthogonal codes, the proposed method significantly amplifies the target peaks in the range-Doppler spectra. In contrast to multistatic radar systems, the proposed approach does not require the sharing of raw radar measurements. Instead, only the weighting matrix is shared among the radars via a wireless channel. Numerical results demonstrate superior detection performance of the proposed constructive interference scheme compared to non-cooperative approaches which treat cross-path signals as mutual interference, underscoring the effectiveness of our collaborative sensing method. This proposed approach significantly benefits automotive radar sensing, especially in detecting weak targets with low radar cross-sections, such as pedestrians and cyclists.


\bibliographystyle{IEEEtran}

\bibliography{refs}

\end{document}